%% file: arxiv_main_SI.tex
\documentclass[aps,pre,twocolumn,superscriptaddress]{revtex4-1}
\usepackage[utf8x]{inputenc}
\usepackage{graphicx}  
\usepackage{dcolumn}   
\usepackage{bm}        
\usepackage{amsmath}   
\usepackage{amssymb}   
\usepackage{siunitx}  

\usepackage[usenames,dvipsnames]{xcolor}

\begin{document}
\widetext


\title{Anisotropic membrane curvature sensing by amphipathic peptides}
\date{\today}


\author{Jordi Gómez-Llobregat}
\email[]{jgomez@turbula.es}
\affiliation{Center for biomembrane research, Department of
  Biochemistry and Biophysics, Stockholm University, SE-106 91
  Stockholm, Sweden.}
\affiliation{Present address: Escola Túrbula, Carretera de
  Mataró, 26 08930 Sant Adrià del Besòs, Barcelona, Spain.}

\author{Federico El\'ias-Wolff}
\email[]{federico.elias.wolff@dbb.su.se}
\affiliation{Center for biomembrane research, Department of
  Biochemistry and Biophysics, Stockholm University, SE-106 91
  Stockholm, Sweden}

\author{Martin Lindén}
\email[]{martin.linden@icm.uu.se}
\affiliation{Department of Cell and Molecular Biology, Uppsala
  University, Box 596, 751 24 Uppsala, Sweden}


\date{\today}

\begin{abstract}
\input{abstract.tex}
\begin{center}
Published version available at \doi{10.1016/j.bpj.2015.11.3512}.
\end{center}
\end{abstract}



\maketitle


\input{maintext.tex}

\input{BPJ_sub2_main.bbl}
\clearpage

\onecolumngrid
\section*{Anisotropic membrane curvature sensing by amphipathic
  peptides\\ -- supporting information.}

\renewcommand{\thesection}{S\arabic{section}}
\setcounter{section}{0}
\renewcommand{\thesubsection}{\Alph{subsection}}
\setcounter{subsection}{0}
\renewcommand{\theequation}{S\arabic{equation}}
\setcounter{equation}{0}
\renewcommand{\thefigure}{S\arabic{figure}}
\setcounter{figure}{0}
\renewcommand{\thetable}{S\arabic{table}}
\setcounter{figure}{0}

\input{supplementarytext_300dpi.tex}

\end{document}

%% file: abstract.tex
Many proteins and peptides have an intrinsic capacity to sense and
induce membrane curvature, and play crucial roles for organizing and
remodelling cell membranes. However, the molecular driving forces
behind these processes are not well understood. Here, we describe a
new approach to study curvature sensing, by simulating the
direction-dependent interactions of single molecules with a buckled
lipid bilayer. We analyse three amphipathic antimicrobial peptides, a
class of membrane-associated molecules that specifically target and
destabilize bacterial membranes, and find qualitatively different
sensing characteristics that would be difficult to resolve with other
methods. These findings provide new insights into the curvature
sensing mechanisms of amphipathic peptides and challenge existing
theories of hydrophobic insertion. Our approach is generally
applicable to a wide range of curvature sensing molecules, and our
results provide strong motivation to develop new experimental methods
to track position and orientation of membrane proteins.

%% file: maintext.tex
\section*{Introduction}
Curvature sensing and generation by membrane proteins and lipids is
ubiquitous in cell biology, for example to maintain highly curved
shapes of organelles, or drive membrane remodelling processes
\cite{Zimmerberg06}. 
Membrane curvature sensing occurs if a molecule's binding energy
depends on the local curvature \cite{Baumgart11}. For proteins,
the presence of multiple conformations with different curvature
preferences can couple protein function to membrane
curvature \cite{tonnesen2014}, with interesting but largely unexplored
biological implications.

Curvature sensing by lipids is often rationalized by a lipid shape
factor, classifying lipids as `cylindrical' or `conical' when they
prefer flat or curved membranes, respectively
\cite{Zimmerberg06,Baumgart11}. Membrane proteins offer a wider range
of sizes, shapes, and anchoring mechanisms \cite{engelman2005}, and
thus potentially more diverse sensing mechanisms. In particular, shape
asymmetry implies that the binding energy depends on the protein
orientation in the membrane plane \cite{Fournier96}, and thus cannot
be a function of only mean and Gaussian curvature, which are
rotationally invariant.  This calls for more complex descriptions, and
one natural extension is to model the binding energy in terms of the
local curvature tensor $C_{ij}$ in a frame rotating with the protein
\cite{Fournier96,Perutkova10,ramakrishnan2010,ramakrishnan2011,ramakrishnan2013,akabori2011,walani2014},
which allows different curvature preferences in different
directions. For example, a preference for longitudinal curvature is
generally associated with proteins that are curved in this direction,
such as BAR domains \cite{peter2004,Blood06}, whereas amphipathic
helices \cite{drin2007} are expected to sense transverse curvature,
since their insertion into the membrane-water interface is
energetically favored if the membrane curves away in the transverse
direction \cite{campelo2008,Campelo14,sodt2014}.

Anisotropic curvature sensing is potentially complex, and theoretical
investigations have demonstrated a wide range of qualitative behavior
in local curvature models
\cite{Fournier96,Perutkova10,ramakrishnan2010,ramakrishnan2011,ramakrishnan2013,akabori2011,walani2014},
but the models have not been rigorously tested. In principle, the
curvature-dependent binding energy landscape $E(C_{ij})$ could be
determined by measuring the Boltzmann distribution of protein
configurations on curved membranes of known shape. However, current
experimental techniques track only protein positions
\cite{zhu2012,sorre2012,Aimon14,shi2015,hsieh2012,ramesh2013,black2014},
and hence orientational information is averaged out. Here, we track
both position and orientation of single molecules, using a
computational approach based on simulated membrane buckling.

\begin{figure}[t]
\begin{center}
\includegraphics[width=85mm]{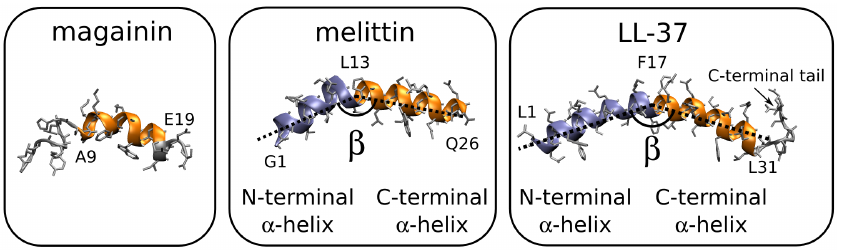}
\end{center}
\caption{\label{Fig1} Structures of magainin \cite{hara2001}, melittin
  \cite{terwilliger1982} and LL-37 \cite{wang2008}. The melittin and
  LL-37 structures contain two $\alpha$-helices that form an angle
  $\beta$ (not the same for both structures).  The $\alpha$-helices
  used in the analysis are colored in blue (N-terminal) and orange
  (C-terminal), with the limiting amino acids labeled on the
  structure. Side chain and non-helical residues are colored in gray.}
\end{figure}

The method is applied to three amphipathic antimicrobial model
peptides: magainin, which is found in the skin of the African clawed
frog \cite{zasloff1987}, melittin, an active component in bee venom
\cite{habermann1972}, and LL-37, a peptide derived from the human
protein cathelicidin which is involved in the innate immune defense
system \cite{gudmundsson1996}. As shown in Fig.~\ref{Fig1}, the
peptides vary in length and shape, and can thus be expected to display
different sensing characteristics. Many antimicrobial peptides are
believed to work by mediating membrane disruption \cite{Melo09}.  The
peptides studied here are thought to mediate the formation of toroidal
membrane pores with a highly curved inner surface partly lined with
lipids
\cite{ludtke1996,yang2001,leontiadou2006,henzler2003,lee2011,sun2015},
although the evidence appears less clear for LL-37 \cite{wang2014}.
The ability to stabilize highly curved membrane structures suggests an
intrinsic preference for curved membranes, as is generally expected
for amphipathic peptides.



Our method uses simulated membrane buckling to sample the
unconstrained interaction of single biomolecules with a range of
membrane curvatures, and extends previous simulation studies of
buckling mechanics \cite{noguchi2011,Hu2013}, curvature-dependent
folding and binding of amphipathic helices \cite{cui2011}, and lipid
partitioning \cite{wang2008}.  We obtain joint distributions of
peptide positions and orientations that yield new biophysical insights
about curvature sensing.  The three model peptides display similar
rotation-averaged curvature preferences but differ in orientational
preferences, which demonstrates the value of directional
information. The asymmetry of the position-orientational distributions
challenges continuum models of amphipathic helices as cylindrical
membrane inclusions \cite{campelo2008,Campelo14}. We speculate that
such asymmetry is important for certain modes of antibacterial
activity, and argue that it might be common also for larger curvature
sensing proteins.  Finally, we discuss the limitations of
characterizing curvature sensing mechanisms from assays with zero
Gaussian curvature, and conclude that this uncertainty affects the
overall binding energy, but not the orientational preferences.  These
results motivate efforts to track positions and orientations of
membrane proteins experimentally, and to develop assays with a broader
range of local curvatures.

\section*{Methods}
\begin{figure*}[t]
\begin{center}
\includegraphics[width=160mm]{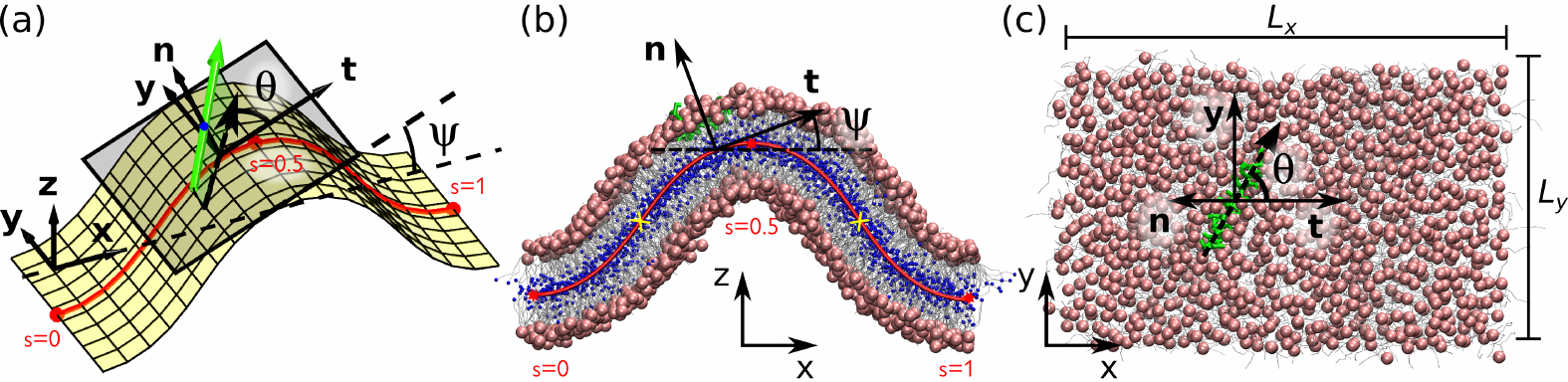}
\end{center}
\caption{\label{Fig2} Buckled simulation and analysis. (a) The
  position $s$ of a peptide is defined by the projection of the
  center-of-mass (blue dot) onto the midplane surface (yellow). The
  in-plane orientation $\theta$ is defined by projecting the peptide
  backbone direction (green arrow, pointing towards the C-terminal
  end) onto the local tangent plane (gray) at $s$. The local tangent
  and normal vectors are indicated by $\textbf{t}$ and $\textbf{n}$,
  respectively. (b) Side and (c) top view of a simulation snapshot
  with peptide position and orientation indicated using the notation
  and local coordinate system in (a).  The system size is $L_x=20.88$
  nm and $L_y=13.05$ nm. The peptide (LL-37 in this case) is shown in
  green, and lipids in gray (tails), light red (phosphate groups) and
  blue (innermost tail beads). The side view (b) also shows the Euler
  buckling profile (red line) fitted to the bilayer mid-plane, and the
  inflection points at $s=0.5\pm 0.25$ (yellow crosses) used to align
  the buckled configurations.  Molecular graphics generated with VMD
  \cite{vmd96}.}
\end{figure*}

To study curvature sensing by single peptides, we simulate their
interactions with a buckled membrane using the coarse-grained Martini
model \cite{Marrink07}, and track their position and orientation, as
shown in Fig.~\ref{Fig2}.  On a microscopic level, curvature sensing
by amphipathic helices is associated with the density and size of
bilayer surface defects \cite{Hatzakis2009,cui2011}, which are well
described by the Martini model \cite{vanni2014}.

\paragraph{Simulation parameters}
We performed molecular dynamics simulations using Gromacs 4.6.1
\cite{Pronk13}, and the coarse-grained Martini force-field with
polarizable water model
\cite{Marrink07,monticelli2008,yesylevskyy2010}, and a relative
dielectric constant of 2.5 (as recommended \cite{yesylevskyy2010}).
We used standard lipid parameters for
1-palmitoyl-2-oleoyl phosphatidylethanolamine (POPE) \cite{Marrink04},
1-palmitoyl-2-oleoyl phosphatidylglycerol (POPG) \cite{Baoukina07},
and peptides \cite{dejong2013}.  The peptide structures for magainin
(PDB ID:1DUM), melittin (PDB ID:2MLT), and LL-37 (PDB ID: 2K6O) were
obtained from the Protein Data Bank, and coarse-grained with the
martinize script provided by the MARTINI developers. Constant
temperature was maintained with the velocity rescaling thermostat
\cite{Bussi07} with a \SI{1.0}{\pico\second} time constant, and
pressure was controlled with the Berendsen barostat
\cite{berendsen1984} using a time constant of \SI{12}{\pico\second}
and a compressibility of \SI{3e-4}{\per\bar}. Peptide (when present),
lipids and solvent were coupled separately to the temperature
bath. Coulomb interactions were modelled with the particle mesh Ewald
method \cite{essmann1995} setting the real-space cut-off to 1.4 nm and
the Fourier grid spacing to 0.12 nm. Lennard-Jones interactions were
shifted to zero between 0.9 and 1.2 nm. A time step of 25 fs was used
in all simulations.

\paragraph{System assembly and membrane buckling}
We assembled and equilibrated three rectangular ($L_x=2L_y$) bilayer
patches of 1024 lipids each, with 70\% POPE and 30\% POPG, solvated
with $\sim 21 000$ coarse-grained water beads and neutralized with
sodium ion beads. POPG is negatively charged, which promotes peptide
binding. These patches were equilibrated for 25 ns in an NPT ensemble
at 300 K and 1 bar, with pressure coupling applied semi-isotropically.

After equilibration, all systems were laterally compressed in the x
direction by a factor $\gamma = (L-L_x)/L = 0.2$, where $L$ is the
linear size of the flat system, and $L_x$ the size of the compressed
simulation box, in the $x$ direction. This was done by scaling all
$x$-coordinates, and the box size $L_x$, by a factor $1-\gamma= 0.8$
at the end of the equilibration run, yielding $L_x = 20.88$, 20.81 and
20.89 nm for the three patches, respectively. After rescaling, the
compressibilities were set to 0 in the $x$ and $y$ directions to keep
the system size constant in those directions for subsequent
simulations. Pressure coupling was then applied anisotropically in the
z direction only. We then performed an energy minimization and a short
equilibration run (25 ns) to let the bilayer buckle.

Next, we added one peptide to each system, using the three independent
patches to create three independent replicas for each peptide. The
peptide was initially placed about 3 nm above the membrane surface,
but quickly attached to the bilayer. After the binding event, we
equilibrated the system for another \SI{5}{\micro\second} before
starting a production run of \SI{15}{\micro\second}, where we
collected data every \SI{5}{\nano\second}.  All peptides remained
essentially parallel to the membrane surface as expected , in
agreement with experimental results for low peptide concentrations
\cite{hara2001,terwilliger1982,lee2013,wang2014}.



\paragraph{Membrane alignment and peptide tracking}
The buckled membrane profile diffuses as a traveling wave the
simulation (movie S1), but curvature sensing by a peptide is reflected
in its distribution relative to the buckled shape. Hence, the buckled
configurations must be aligned in order to extract useful information.
To do this, we fit the $xz$-profile of the membrane by the ground
state of the Helfrich model with periodic boundary conditions, which
is one of the Euler buckling profiles of an elastic beam
\cite{noguchi2011,Hu2013}. This shape depends only on the
dimensionless buckling parameter $\gamma$ ($\gamma=0$ is the flat
state). Hence, if we compute the shape for some reference system, the
general case can be obtained by shifting and scaling. We chose $L_x=1$
as reference, and write the buckling profile as a parametric curve
$x=s+\xi(s,\gamma),z=\zeta(s,\gamma)$, parameterized by a normalized
arclength coordinate $0<s<1$ (the absolute arclength is given by
$sL=sL_x(1-\gamma)^{-1}$).


For fast evaluation, we expanded $\xi(s,\gamma)$ and $\zeta(s,\gamma)$
in truncated Fourier series in $s$, and created look-up tables for
Fourier coefficients vs.~$\gamma$. We defined $s$ to give the curve
$z(x)$ a maximum at $s=0.5$, minima at $s=0,1$, and inflection points
at $s=0.5\pm 0.25$, and aligned the buckled shapes by fitting the
bilayer in each frame to the buckling profile and aligning the
inflection points (Fig.~\ref{Fig2}, movies S2-S3). Specifically, we
fit the rescaled buckling profile to the innermost tail beads of all
lipids in each frame using least-squares in the $x$ and $z$
directions, i.e., minimizing
\begin{equation}
\sum_i(x_0+L_x\big(s+\xi(s_i,\gamma)\big)-x_i)^2
        +(z_0+L_x\zeta(s_i,\gamma)-z_i)^2
\end{equation}
with respect to $\gamma$, the translations $x_0,z_0$, and the
normalized arc-length coordinates $s_i$ of each bead ($x_i,z_i$ are
bead positions). The time-averaged bilayer shape, after alignment,
agrees well with the theoretical buckled shape (Fig.~\ref{Fig3}a).

The normalized arclength coordinate $s$ of the peptide was computed by
projecting the peptide center of mass onto the buckled profile fitted
to the membrane midplane.  The in-plane orientation $\theta$ was then
computed by fitting a line through the backbone particles of the
$\alpha$-helical part of the peptide, projecting it onto the tangent
plane at $s$, and computing the in-plane angle to the tangent vector
$\mathbf{t}$ (see Fig.~\ref{Fig2}a).  The local curvature at $s$, in
the tangent direction of the buckled shape, is given by
  \begin{equation}
C(s)=\frac{1-\gamma}{L_x}\frac{d\psi}{ds},     
\end{equation}
where $\psi$ is the bilayer mid-plane tangent angle of
Fig.~\ref{Fig2}a,b \cite{kreyszig1991}. (Note that the opposite sign
convention is also common \cite{Zimmerberg06}) . In the theoretical
analysis, we neglect small shape and area fluctuations
($\mathrm{std}(\gamma)\approx 0.005$) and use the nominal value
$\gamma=0.2$.

\paragraph{Fitting}
We used least-squares routines in MATLAB (MathWorks, Natick, MA) to
fit the Boltzmann distributions $e^{-E(s_i,\theta_i)}/Z$ of the $E_C$
(Eq.~\ref{eq:EH}) and $E_2$ (Table \ref{E2par}) models to
$(s,\theta)$-histograms built from the aggregated data with 50 bins
for each coordinate. Both data and model histograms were normalized
numerically.  Error bars in Fig.~\ref{Fig4}d are boot-strap standard
deviations from 1000 bootstrap realizations, using blocks of length
100 (500 ns) as the elementary data unit for resampling
\cite{kunsch1989}.

\section*{Results}
\begin{figure}[h!]
\begin{center}
\includegraphics[width=85mm]{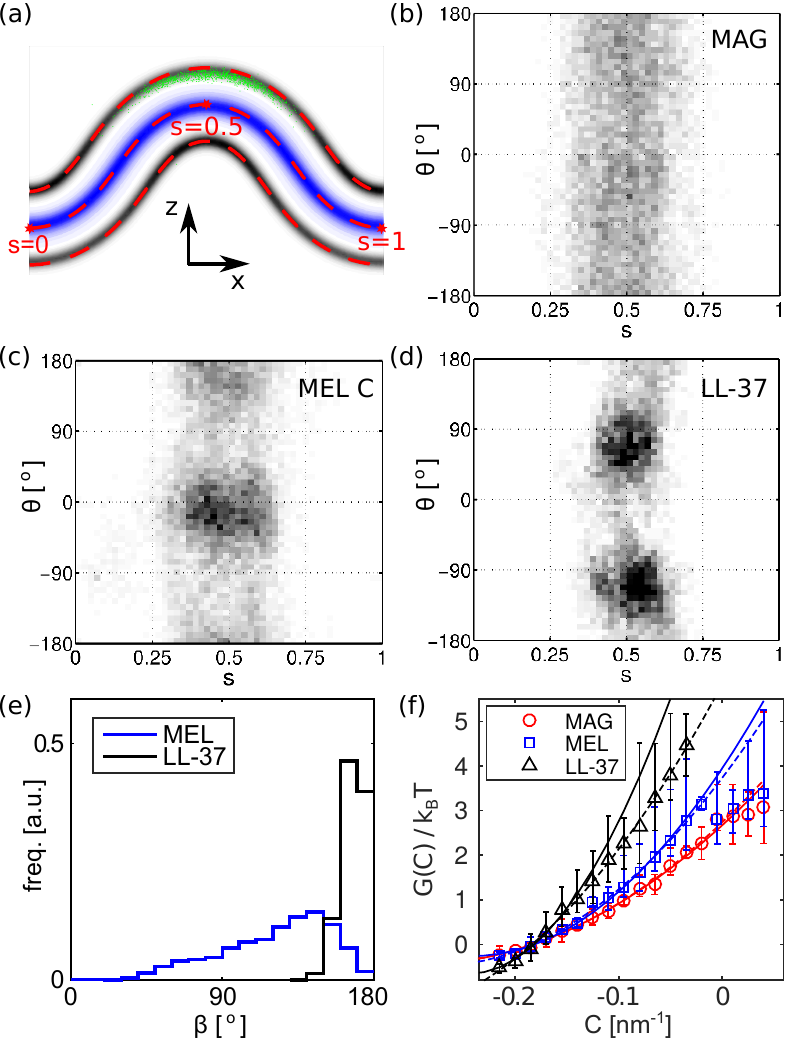}
\end{center}
\caption{\label{Fig3} Distributions of peptide positions and
  orientations in the buckled bilayer. (a) Average buckled shape in
  terms of densities of inner lipid tail beads (blue) and phosphate
  groups (gray) after alignment, for one production run with
  LL-37. Green dots show representative peptide center-of-mass
  positions. Dashed red lines indicate the average fitted mid-plane
  $\pm 2.15$ nm offsets in the normal direction. (b-d) Aggregated
  $(s,\theta)$-histograms for (b) magainin, (c) melittin (using the
  orientation of the C-terminal helix), and (d) LL-37. (e)
  Distributions of internal angle $\beta$ (see Fig.~\ref{Fig1}) for
  melittin and LL-37. (f) Orientation-averaged binding free energy
  vs.~curvature at the peptide center-of-mass
  (Eq.~\eqref{eq:GChistogram}) for the three peptides. Error bars show
  max and min values from three independent simulations. Dashed lines
  are guides to the eye (fits to quadratic curves), and solid lines
  are results for the $E_C$ model (Eq.~\eqref{eq:EH}) using the fit
  parameters in Fig.~\ref{Fig4}.}
\end{figure}

\paragraph{Preferred curvature and orientations}
We simulated single peptides interacting with a buckled bilayer, using
three independent production runs of 15 $\mu s$ for each peptide, and
tracked their normalized arc-length coordinates $s\in[0,1]$ and
in-plane orientations $\theta$ (Fig.~\ref{Fig2}a).  Aggregated
$(s,\theta)$-histograms are shown in Fig.~\ref{Fig3}b-d, and
convergence is discussed in Sec.~\ref{Stext:replicas}.

All three peptides prefer the concave high curvature regions with a
maximum at $s=0.5$, as expected for hydrophobic insertion mechanisms
\cite{campelo2008,Campelo14,sodt2014,Hatzakis09,cui2011,vanni2014}.
Regarding the angle distributions, the three peptides behave
differently. Magainin displays a rather uniform angle distribution,
probably because its short $\alpha$-helical segment creates a fairly
symmetric insertion footprint.  For melittin, the joint between the N-
and C-terminal helices appears very flexible, resulting in a broad
distribution of the internal angle $\beta$ (Fig.~\ref{Fig3}e).  Both
helices prefer directions nearly parallel to the x-axis, the direction
of maximum curvature, but the preference is stronger and slightly
offset ($\theta_{max}\approx -15^\circ,165^\circ$) for the C-terminal
helix shown in Fig.~\ref{Fig3}c, while the N-terminal helix is more
symmetrically oriented (Fig.~\ref{Sfig:MELreplicas}).

LL-37 maintains a linear structure, and its $\theta$-distribution
displays two sharp maxima near $\theta=70^\circ$ and
$\theta=-110^\circ$ (Fig.~\ref{Fig3}d). This is remarkable since, by
reflection symmetry around $s=0.5$, the curvatures in those directions
are the same as along $-70^\circ$ and $110^\circ$, orientations that
are clearly not preferred.  As we will argue below, this can be
understood as curvature sensing along directions different from that
of the peptide itself.  These sensing directions adopt
$\theta=0,90^\circ$, and thus map onto themselves under reflection.
Notably, none of the peptides orient directly along the flat direction
$\theta=90^\circ$ as commonly assumed in mechanical models
\cite{campelo2008,Campelo14}.

\paragraph{Orientation-averaged binding free energy}
Next, we look at the orientation-averaged binding free energy,
corresponding to the curvature-dependent enrichment measured in many
\textit{in vitro} assays
\cite{zhu2012,sorre2012,Aimon14,shi2015,hsieh2012,ramesh2013,black2014}.
To extract the curvature dependence of the binding energy, we analyse
center-of-mass positions along the buckled shape. These should follow
a Boltzmann distribution, proportional to $e^{-G(s)}$, where $G(s)$ is
the orientation-averaged binding free energy in units of $k_BT$. We
model this as depending on the local curvature only, and hence set
$G(s)=G(C(s))$, and extract $G(C)$ from curvature histograms, weighted
according to the change-of-variable transformation that relates the
density of curvatures, $\rho(C)$, to the density of positions
$\rho(s)$. Indeed, dropping normalization constants, we have
\begin{equation}
\rho_s(s)ds
\propto e^{-G(C(s))} ds 
\propto e^{-G(C)}|dC/ds|^{-1}dC
\propto \rho_C(C)dC,
\end{equation}
from which it follows that
\begin{equation}\label{eq:GChistogram}
G(C)=-\ln\big(\rho_C(C)|dC/ds|\big)+const.
\end{equation}
The weights $|dC/ds|$ can be understood as compensating for the fact
that not all curvatures have equal arclength footprints along the
buckled profile. To estimate $G(C)$, we estimated $\rho(C)$ using a
simple histogram, and the weights as the mean of $|dC/ds|$ for all
contributions to each bin.

Fig.~\ref{Fig3}f shows the binding free energy profiles $G(C)$ for the
different peptides, which are more similar than the
$(\theta,s)$-distributions, and well fit by quadratic curves.  Note
that Eq.~\eqref{eq:GChistogram} does not yield absolute binding
energies of the peptides, and the $G(C)$ curves are instead offset
vertically for easy visualization. Experimental binding free energies
of these peptides to flat membranes with anionic lipids range from -15
to -10 $k_BT$ \cite{he2013}.

\paragraph{Quantitative models}
We now turn to quantitative models of the peptides' curvature sensing.
As described in the introduction, we model the binding energy of a
peptide as a function of the local curvature tensor in a frame
rotating with the peptide, and treat the bilayer itself as having
fixed shape and thus a fixed deformation energy which we neglect.
Generally, if the principal curvatures and directions are $c_{1,2}$
and $\vec{e}_{1,2}$, the curvature tensor, or second fundamental form,
in a frame rotated by an in-plane angle $\theta$ relative to
$\vec{e}_1$, is given by
\begin{equation}\label{eqCij}
C_{ij}
=\left[\begin{array}{cc} H\!\!+\!\!D\cos2\theta& D\sin2\theta\\
D\sin2\theta&H\!\!-\!\!D\cos2\theta\end{array}\right]
=\left[\begin{array}{cc}C_\parallel&C_X\\C_X&C_\perp\end{array}\right],
\end{equation}
where $H=(c_1+c_2)/2$ and $D=(c_1-c_2)/2$ are the mean and deviatoric
curvatures, and the Gaussian curvature is given by
$K=c_1c_2=C_\parallel C_\perp-C_X^2$. Note the symmetry under
rotations by $180^\circ$, since the curvature of a line is the same in
both directions. For the buckled surface, $c_1=C(s),\,c_2=0$ (and
hence $K=0$, $H=D=C(s)/2$), $\vec{e}_{1}=\mathbf{t}$,
$\vec{e}_2=\mathbf{y}$. As shown in Fig.~\ref{Fig2}, we define the
rotating frame using the peptide's center of mass and the direction of
the $\alpha$-helical parts, and thus $\theta$ is the peptide in-plane
orientation, and $\parallel,\perp$ denote the longitudinal ($\theta$)
and transverse ($\theta+90^\circ$) directions.

The simplest models are linear in $C_{ij}$, but can be ruled out since
they cannot reproduce the convex binding free energies in
Fig.~\ref{Fig3}f. To see this, we write a general linear model in the
form $E_1=aH+bD\cos(2(\theta-\alpha))$ \cite{Fournier96}, and
integrate out the angular dependence to get
\begin{equation}\label{eq:G1}
G_1=-\ln\int_0^{2\pi}e^{-E_1}d\theta
        =aH-\ln I_0(bD)+const.
\end{equation}
Since $H=D=C(s)/2$ on the buckled surface, and the modified Bessel
function $I_0$ is convex, $G_1$ will be either downward convex (if
$b\ne 0$) or linear and direction insensitive (when $b\to 0$), in
disagreement with Fig.~\ref{Fig3}.

Moving on to quadratic terms, Akabori and Santangelo \cite{akabori2011}
explored a model of the form
\begin{equation}\label{eq:EAS}
  E_X=\!\frac{k_{\parallel}}{2}(C_\parallel\!-C_{\parallel0})^2
  \!\!+k_X(C_X-C_{X0})^2 
  \!\!+\frac{k_{\perp}}{2}(C_\perp\!-C_{\perp0})^2,
\end{equation}
where $C_{\parallel0}$, $C_{X0}$ and $C_{\perp0}$ are preferred
curvatures.  Further simplifications $k_X$=0 and $k_X$=$k_\perp$=0
have also been
studied \cite{Perutkova10,ramakrishnan2010,ramakrishnan2011,ramakrishnan2013}.
While these models can all display non-trivial behavior, $E_X$ is
not the most general quadratic model, which would include all 9 linear
and quadratic combinations of the three independent curvature tensor
components.  In particular, $E_X$ does not contain a simple preferred
mean curvature as a special case, because $H=(C_\parallel+C_\perp)/2$,
and hence $(H-H_0)^2$ contains a term $C_\parallel C_\perp$ which is
absent in Eq.~\eqref{eq:EAS}.

However, the general quadratic model is not identifiable on surfaces
with only one non-zero principal curvature. This is because the
Gaussian curvature $K$ is zero, and hence the model can only be
specified up to a term proportional to $K$. Also, $E_X$ can then be
made to behave as a mean curvature sensor, since all angular
dependence cancels if $k_\parallel=k_\perp=k_X$,
$C_{\parallel0}=C_{\perp0}$, and $C_{X0}=0$.  These limitations apply
to our buckled surface, as well as to tubular and plane-wave
geometries used experimentally
\cite{sorre2012,zhu2012,hsieh2012,ramesh2013,black2014}.  A curvature
sensing mechanism therefore cannot be completely characterized using
such surfaces, but some conclusions can be drawn.

\begin{figure}
\begin{center}
\includegraphics[width=85mm]{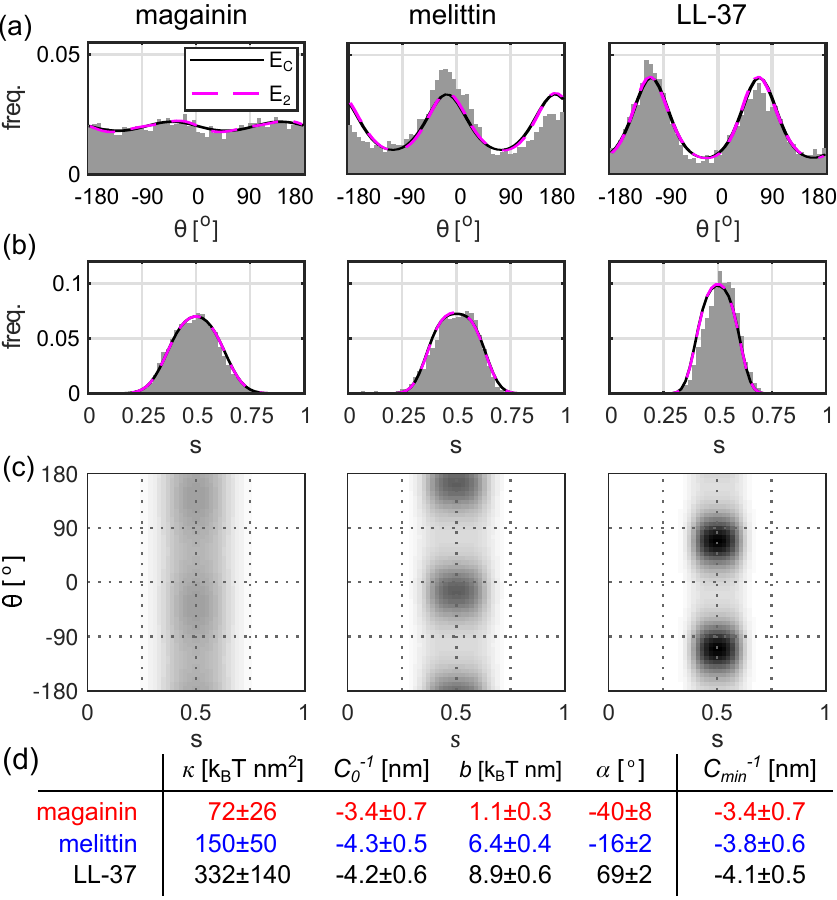}
\end{center} 
  \caption{\label{Fig4} Fitting quadratic models to data.  (a,b)
    marginal position and angle distributions, showing data (gray) and
    nearly identical curves from the $E_C$ and general quadratic model
    ($E_2$, see Table \ref{E2par}). (c) $(s,\theta)$-distributions for
    the $E_C$ model. (d) $E_C$ fit parameters $\pm$ bootstrap SEM
    \cite{kunsch1989} due to finite sampling (see
    Sec.~\ref{Stext:replicas}), with $\alpha$ indicating the preferred
    orientation. $C_{min}$ is the preferred curvature, from minimizing
    Eq.~\ref{eq:GC}.}
\end{figure}

In particular, setting $k_X=0$ in Eq.~\eqref{eq:EAS} yields an
intuitive model with curvature sensing only along the longitudinal and
transverse directions
\cite{Perutkova10,ramakrishnan2010,ramakrishnan2011,ramakrishnan2013}.
From Eq.~\eqref{eqCij}, this means angular dependence only in the form
$\cos 2\theta$, which is symmetric around $\theta=0$,
$\pm\frac{\pi}{2}$, and $\pm\pi$. However, the orientational
distributions in Fig.~\ref{Fig4}a do not display this symmetry,
although the statistics is not quite clear in the case of melittin
(see Fig.~\ref{Sfig:MELreplicas}). Apparently, the curvature sensing
directions are not generally aligned with the actual helices.  This
resembles results for $\alpha$-synuclein, where peptides and induced
membrane deformations appear similarly misaligned \cite{Braun12}.  A
simple quadratic model incorporating these observations is
\begin{equation}\label{eq:EH}
E_C=\frac{\kappa}{2}(2H-C_0)^2+bD\cos\big(2(\theta-\alpha)\big)
+\kappa_GK,
\end{equation}
where the Gaussian curvature coefficient $\kappa_G$ is unidentifiable
since $K=0$ in our data. As shown in Fig.~\ref{Fig4}, $E_C$ describes
all peptides reasonably well, and using the full quadratic model does
not significantly improve the fit.

To better understand the physical meaning of this model, we explore
some alternative formulations. First, using Eq.~\eqref{eqCij} to trade
$H,D$ for the $C_{ij}$, and rearranging the terms, we find an
equivalent formulation that resembles the $E_X$ model,
\begin{multline}
  E_C'=
  \frac{\kappa}{2}\big(C_\parallel-C_0+\frac{b}{2\kappa}\cos2\alpha\big)^2
  +\kappa\big(C_X+\frac{b}{2\kappa}\sin2\alpha\big)^2
  \\ +\frac{\kappa}{2}\big(C_\perp-C_0-\frac{b}{2\kappa}\cos2\alpha\big)^2
  +(\kappa+\kappa_G)K.
\end{multline}
Continuing, we can rotate the basis attached to the peptide by
$\alpha$, and thus generate a transformed curvature tensor with
elements $C_{ij}^{(\alpha)}(\theta)=C_{ij}(\theta+\alpha)$ satisfying
\begin{equation}
C_{\parallel}^{(\alpha)}+C_{\perp}^{(\alpha)}=2H,\quad
C_{\parallel}^{(\alpha)}-C_{\perp}^{(\alpha)}
        =2D\cos\big(2(\theta-\alpha)\big).
\end{equation}
In this basis, there is an $E_X$-like equivalent model that lacks
'off-diagonal' elements,
\begin{equation}\label{eq:ECpp}
  E_C''=
  \frac{\kappa}{2}\big(C_{\parallel}^{(\alpha)}-C_0+\frac{b}{2\kappa}\big)^2
  +\frac{\kappa}{2}\big(C_{\perp}^{(\alpha)}-C_0-\frac{b}{2\kappa}\big)^2
  +(\kappa+\kappa_G)K,
\end{equation}
i.e., sensing curvature along two orthogonal directions that are
rotated by an angle $\alpha$ with respect to the peptide
backbone. Note that since Gaussian curvature is rotationally
invariant, the unidentifiable Gaussian curvature term only affects the
overall affinity to membranes with Gaussian curvature, and not the
orientational preferences of the peptides.

\begin{figure}
  \includegraphics[width=85mm]{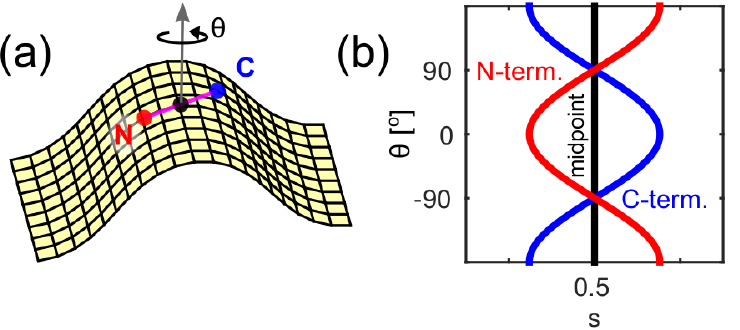}
  \caption{\label{Fig5} Curvature sensing site and
    $(s,\theta)$-correlations. (a) A freely rotating peptide whose
    midpoint (black) is fixed at $s=0$. (b) Resulting correlations
    $s_{C,N}\propto \pm\cos\theta$ for the N- and C-terminal ends
    (red,blue).}
\end{figure}

As a consistency check, we integrated out $\theta$ from $E_C$.
Proceeding as for $G_1$ in Eq.~\eqref{eq:G1} and setting $H=D=C(s)/2$,
$K=0$, we get
\begin{equation}\label{eq:GC}
  G_C=-\ln\int_0^{2\pi}d\theta e^{-E_C}
  =\frac{\kappa}{2}(C-C_0)^2-\ln I_0\big(bC/2\big),
\end{equation}
which we compare with $G(C)$ in Fig.~\ref{Fig3}f using the parameters
of Fig.~\ref{Fig4}d.  Magainin and melittin shows good agreement, but
not LL-37, whose $(s,\theta)$-distribution (Fig.~\ref{Fig3}d) is also
less symmetric around $s=0.5$ than expected from the symmetry of the
buckled shape.  A simple explanation is that the effective ``sensing
site'' does not coincide with the center of mass used to define $s$.
This is illustrated in Fig.~\ref{Fig5} by a hypothetical peptide which
is fixed at $s=0.5$ but free to rotate. As a result, the N-terminal
end shows $(s,\theta)$-correlations resembling those seen for the
LL-37 center of mass, indicating that its ``sensing site'' is located
in the C-terminal part. Numerical experiments in Sec.~\ref{LL37} agree
qualitatively with this geometric argument, and both symmetry and
consistency improves when tracking the LL-37 C-terminal helix instead
(but the fit parameters do not change significantly).

%


\section*{Discussion} 
We describe a simulation approach to study membrane curvature sensing
by tracking positions and orientations of single molecules interacting
with a buckled lipid bilayer. This approach is widely applicable, and
the utility of angular information is obvious from the observation
that the three peptides show distinct orientational distributions, but
very similar orientation-averaged binding energy curves
(Fig.~\ref{Fig3}).

Our data is well described by modelling the binding energy in terms of
local curvatures, yielding more complex models than commonly used to
fit orientation-averaged data
\cite{zhu2012,sorre2012,Aimon14,shi2015}, and also less symmetric than
some theoretical suggestions
\cite{Perutkova10,ramakrishnan2010,ramakrishnan2011,ramakrishnan2013}.
The observed asymmetry also seems difficult to reconcile with
continuum elasticity models of hydrophobic insertion in terms of
cylindrical membrane inclusions
\cite{campelo2008,Campelo14}. Recently, continuum elasticity models
were found to underpredict the induced curvature of a hydrophobic
insertion compared to atomistic simulations \cite{sodt2014}. Our data
shows an additional qualitative effect of molecular detail, which we
believe reflect the fact that the mirror symmetry of cylinder-shaped
inclusions is absent from the peptide structures. Instead, our data
can be described in terms of curvature sensing directions that are not
aligned with the inserted $\alpha$-helices. Since amphipathic helices
are common curvature sensing motifs \cite{drin2007} and mirror
symmetry is generally absent also in multimeric proteins
\cite{goodsell00}, such asymmetric sensing might be common.

These results should motivate efforts to track the position and
orientation of membrane proteins experimentally, for example using
polarization-based optical techniques \cite{rosenberg2005} or electron
microscopy \cite{davies2012}.  It would also be valuable to vary mean
and Gaussian curvatures independently in order to probe Gaussian
curvature sensing, for example by extending supported bilayer assays
with plane-wave surfaces\cite{hsieh2012} to shapes with non-zero
Gaussian curvature. Another possibility might be to combine assays
with cylindrical geometries ($K=0$), such as plane waves
\cite{hsieh2012} or membrane
tethers\cite{zhu2012,sorre2012,Aimon14,shi2015,ramesh2013} with
spherical geometries ($K=H^2$) such as vesicles
\cite{Hatzakis09,tonnesen2014} or deposited nanoparticles
\cite{black2014}.

An interesting aspect of the $E_C$ model is that it predicts a free
energy minimum, i.e., a preferred curvature, at least when $K=0$
(Eq.~\ref{eq:GC}). The preferred curvature radii $C_\text{min}^{-1}$
of our peptides, listed in Fig.~\ref{Fig4}c, are well above the
monolayer thickness of about 2.2 nm (Fig.~\ref{Fig3}a) where the
bilayer folds back on itself, but below the lowest radius in our
simulations (about 4.5 nm), meaning that this prediction is somewhat
speculative, since higher order terms might become important at very
high curvatures.  On a molecular level, curvature sensing by
amphipathic peptides is thought to reflect an affinity for packing
defects in the membrane-water interface
\cite{Hatzakis09,cui2011,vanni2014}. It is not clear that this
mechanism predicts a preferred curvature at all. Testing this seems
like an interesting question for future work.

Our results also have biophysical implications.  At high
concentrations, the three peptides are thought to mediate the
formation of membrane pores with highly curved inner surfaces
\cite{ludtke1996,yang2001,leontiadou2006,henzler2003,lee2011,sun2015}. The
orientational preferences we see in single peptides are consistent
with atomistic \cite{leontiadou2006} and coarse-grained \cite{sun2015}
simulations of multi-peptide pores.  In particular, the asymmetric
curvature preference of LL-37 should help select for a single
handedness of the resulting tilted pore structure \cite{sun2015},
which might facilitate pore formation by reducing frustration. This
mechanism may represent a general way for membrane proteins to induce
a particular orientation or handedness in patterns on curved surfaces
\cite{mim2012}.
\medskip\\
\paragraph{Acknowledgments}
We thank Astrid Gräslund, Oksana V.~Manyuhina, Christoph
A.~Haselwandter, and two anonymous reviewers for helpful comments and
discussions.  Simulations were performed on resources provided by the
Swedish National Infrastructure for Computing (SNIC) at the National
Supercomputer Centre (NSC) and the High Performance Computing Center
North (HPC2N). Financial support from the Wenner-Gren Foundations and
the Swedish Foundation for Strategic Research (SSF) via the Center for
Biomembrane Research are gratefully acknowledged.
\paragraph{Author contributions}
JG and ML designed research. JG and FEW performed research. JG and ML
analysed data. JG, FEW, and ML wrote the paper.

%% file: BPJ_sub2_main.bbl
\providecommand{\noopsort}[1]{}\providecommand{\singleletter}[1]{#1}%

%% file: supplementarytext_300dpi.tex
\section{Convergence and individual replicas}\label{Stext:replicas}
\begin{figure*}[b]
  \begin{center}
    \includegraphics[width=110mm]{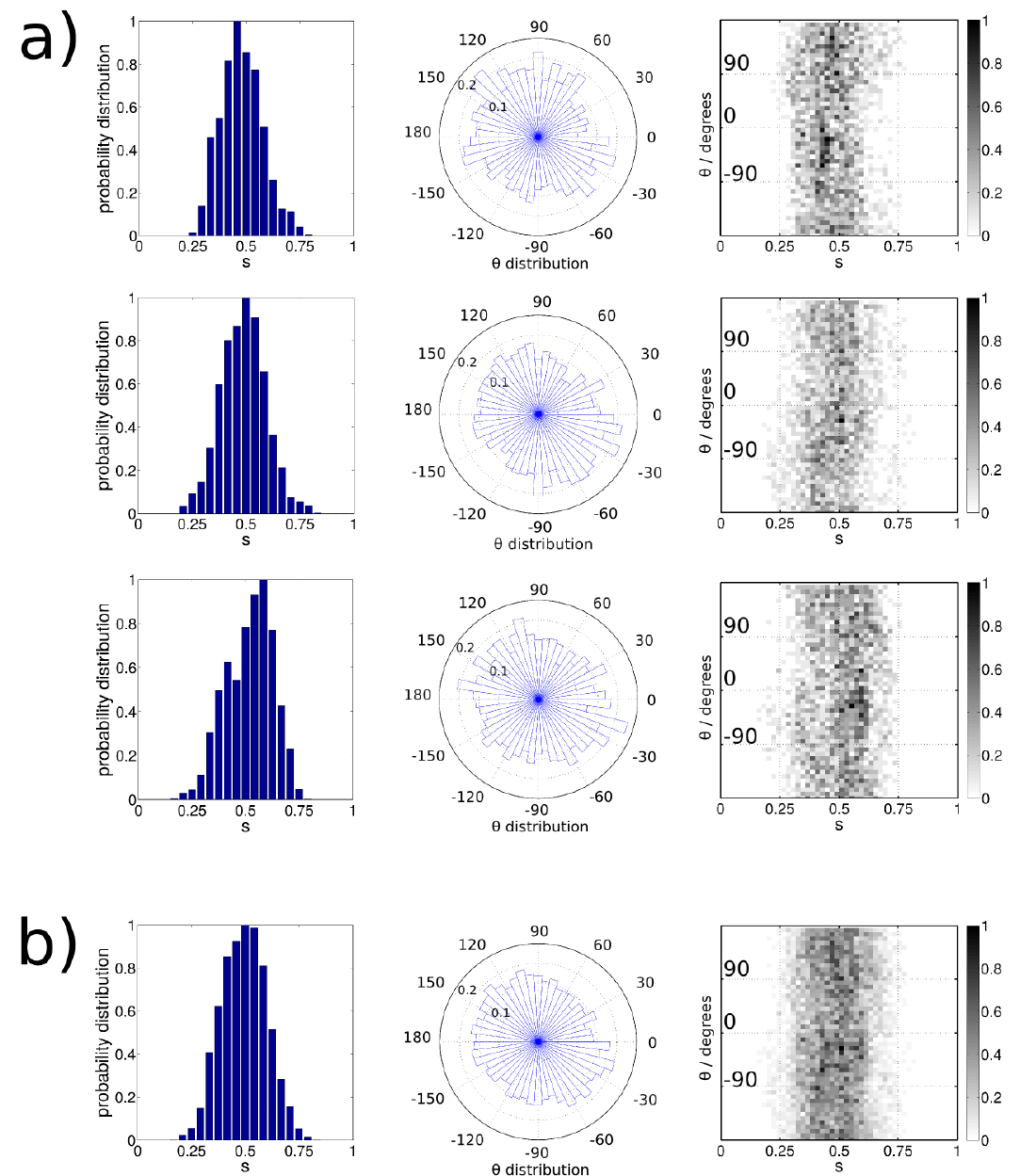}
  \end{center}
  \caption{Results for magainin, from (a) three independent production
    runs, and (b) aggregated.}\label{Sfig:MAGreplicas}
\end{figure*}

\begin{figure*}[h]
  \begin{center}
    \includegraphics[width=170mm]{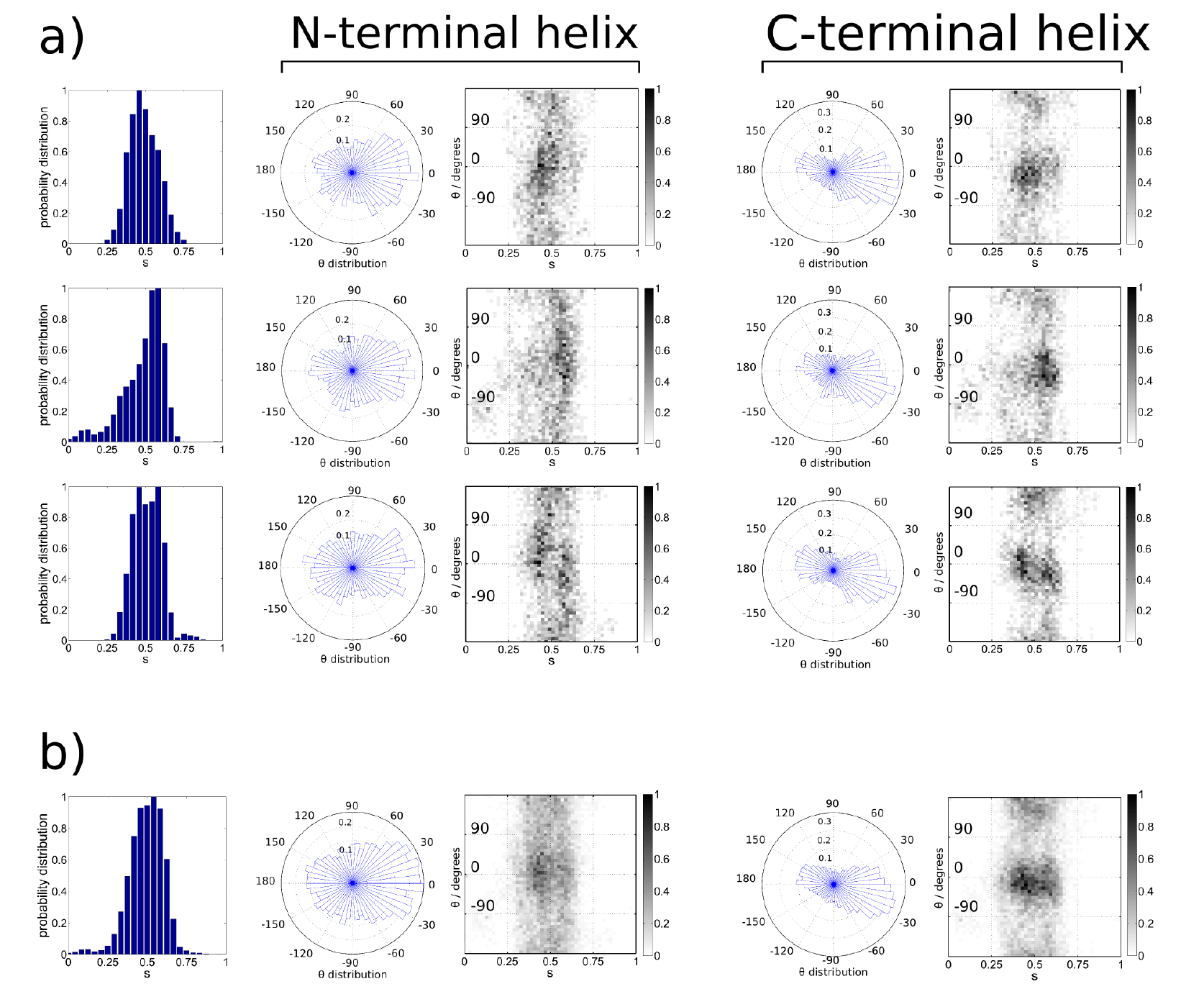}
  \end{center}
  \caption{Results for melittin, from (a) the three independent
    production runs, and (b) aggregated. Both N- and C-terminal
    results are shown. }\label{Sfig:MELreplicas}
\end{figure*}

\begin{figure*}[h]
  \begin{center}
    \includegraphics[width=110mm]{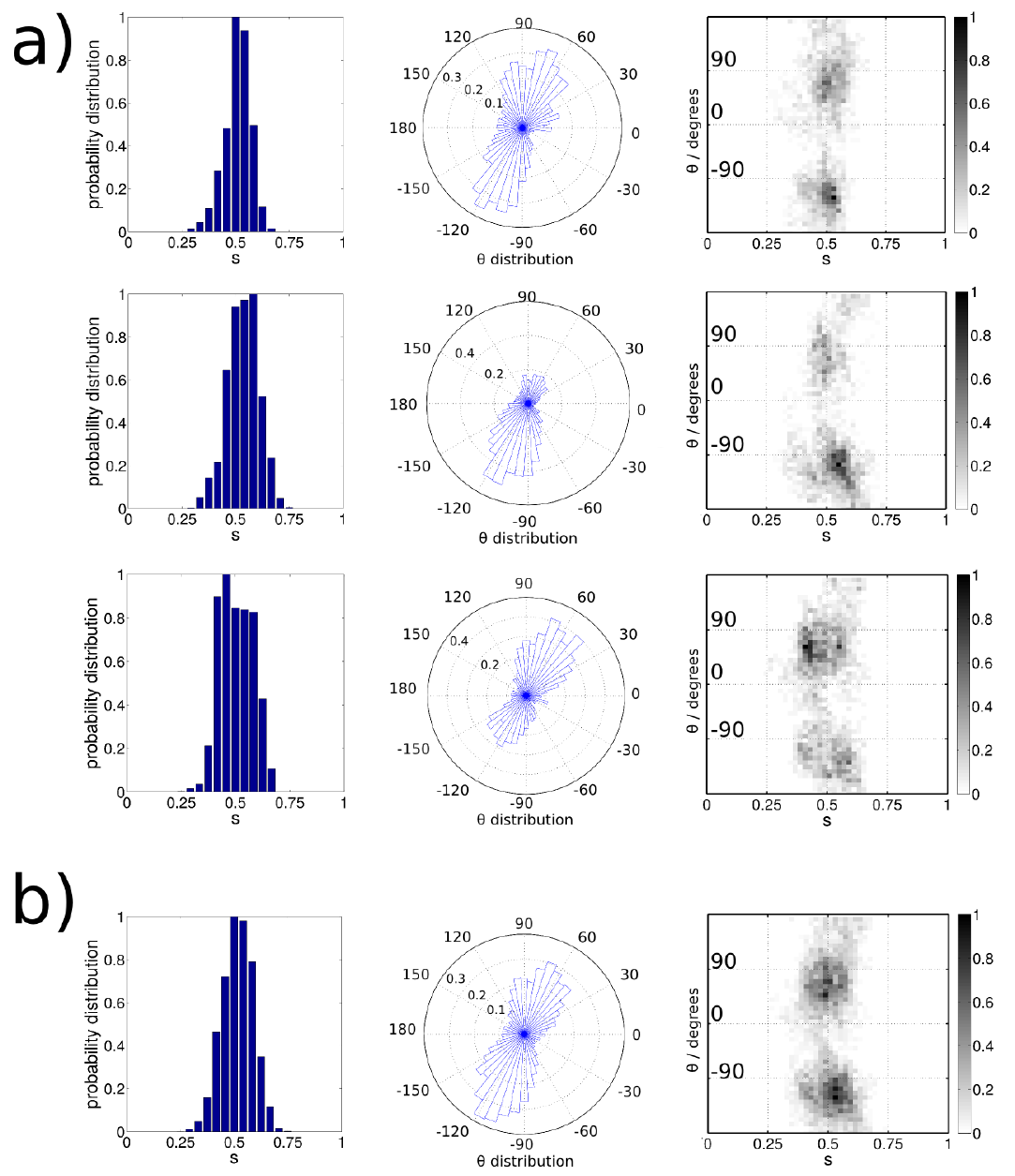}
  \end{center}
  \caption{Results for LL-37, from (a) three independent production
    runs, and (b) aggregated.}\label{Sfig:LL37replicas}
\end{figure*}

Simulations of proteins interacting with mixed bilayers can be
challenging to converge due to slow lipid diffusion and long-lived
protein-lipid interactions \cite{ge2014}. For this reason, we run
three independent replicas rather than one long simulation for each
peptide, and use them as a simple control of the robustness of our
conclusions.  Figures \ref{Sfig:MAGreplicas}-\ref{Sfig:LL37replicas}
show histograms of center-of-mass positions, orientations, and joint
positions-orientations of both the three individual production runs
for each peptide, as well as aggregated histograms. In the case of
melittin (Fig.~\ref{Sfig:MELreplicas}), orientations of both the N-
and C-terminal helices are shown.

While the results for individual trajectories are obviously noisier
than the aggregated statistics, it is clear that the same qualitative
features are present in all replicas. In particular, two
well-separated orientational states of melittin and LL-37 are clearly
visible (albeit not equally populated) in all trajectories, strongly
indicating that our simulations are long enough to capture the major
low-energy states of these systems.  However, the sampling is still
limited enough to induce significant statistical uncertainty in fit
parameters, as seen Fig.~\ref{Fig4}d.

\begin{table}[b]
  \caption{ \label{E2par}Fit parameters (fit$\pm$ bootstrap std.) for
    the $E_2$ model for the curves shown in Fig.~\ref{Fig4}, rounded
    to two significant digits, in appropriate units of $k_BT$ and
    nm. This model is given by $E_2=\frac{a_1}{2}(C_\parallel-a_2)^2
    +a_3(C_X-a_4)^2+\frac{a_5}{2}(C_\perp-a_6)^2
    +a_7C_X(C_\parallel+C_\perp)+a_8C_X(C_\parallel-C_\perp),$ i.e.,
    with a $C_\parallel C_\perp$-term omitted for identifiability.}
  \begin{center}
        \begin{tabular}{|l|c|c|c|c|}
          \hline
          & $a_1$& $a_2$& $a_3$& $a_4$\\
          \hline
          MAG & 
          $ 71\pm 30$ & $-0.30 \pm 0.26 $&
          $ 73\pm 28$ & $ 0.024\pm 0.028$\\
          MEL & 
          $170\pm 52$ & $-0.23 \pm 0.03 $& 
          $130\pm 52$ & $0.002 \pm 0.02 $\\
          LL-37    &
          $300\pm130$ & $-0.24\pm0.07$& 
          $300\pm140$ & $-0.02\pm0.03$\\
          \hline
          \hline
          & $a_5$& $a_6$& $a_7$& $a_8$\\
          \hline
          MAG & 
          $ 68\pm 28$ & $-0.30 \pm 0.27 $&
          $-24\pm 19$ & $5.3\pm3.3$\\
          MEL & 
          $ 90\pm 55$ & $-0.3  \pm 1.3  $&
          $ 27\pm 28$ & $2.4\pm4.2$\\
          LL-37&
          $310\pm150$ & $-0.25\pm0.73$&
          $68 \pm 59$ & $ 1 \pm 7$\\
          \hline
        \end{tabular}
  \end{center}
\end{table}
\clearpage
\section{Location of the curvature sensing site on LL-37}\label{LL37}
LL-37 shows indications of asymmetry around $s=0.5$ that is
incompatible with the symmetry of the curvature tensor elements
(Fig.~\ref{Fig3}d), and the fitted $E_C$ model is also less consistent
with the orientation averaged binding energy (Fig.~\ref{Fig3}e) than
the other peptides.  Here, we explore the hypothesis that these
effects are caused by using the center-of-mass of the peptide for
defining the position $s$, which might be inappropriate if the
sensitivity is unequally distributed along the peptide.  Our rationale
for this hypothesis is that a correlation between position and
orientation, as indicated in the LL-37 data in Fig.~\ref{Fig3}d might
come about if the effective curvature sensing site is different than
the center-of-mass which we tracked to extract that data, as sketched
in Fig.~\ref{Fig5}.

\begin{figure*}[h]
  \includegraphics{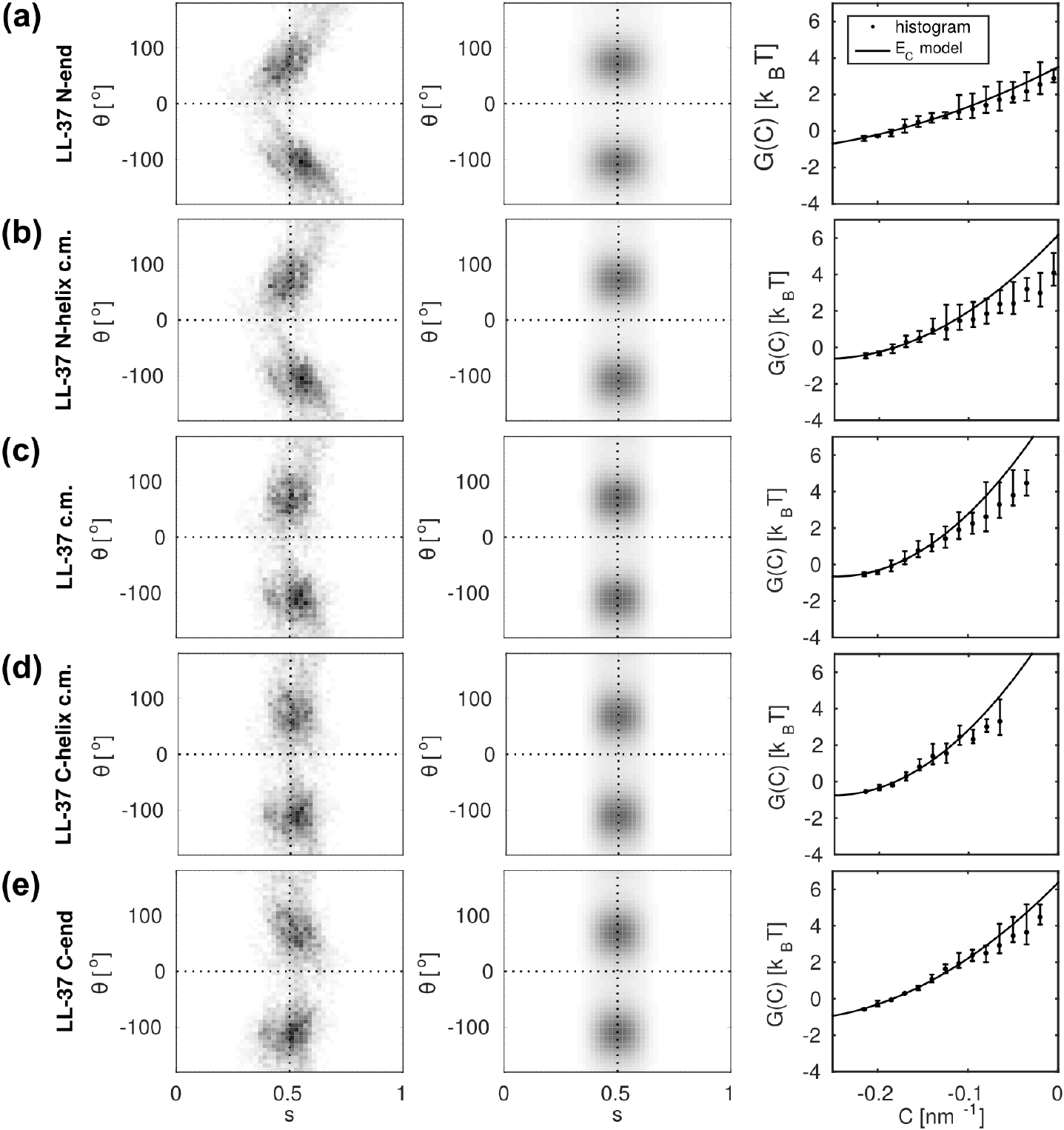}
  \caption{Analysis of LL-37 using different definitions of $s$ and
    $\theta$, namely (a) the first residue and orientation of the
    N-terminal helix, (b) the center-of-mass and orientation of the
    N-terminal helix, (c) the center-of-mass and orientation of the
    whole peptide (same as shown in the main text), (d) the
    center-of-mass and orientation of the C-terminal helix, and (e)
    the last residue and orientation of the C-terminal helix.
    relevant curvature sensing site. The columns show (left) the
    ($s,\theta$)-histogram, (mid) a fit of the $E_C$ model, and
    (right) the orientation-averaged binding free energy, obtained
    from the model fit (line) or using weighted histograms,
    Eq.~\ref{eq:GChistogram}, (dots) with error bars as in
    Fig.~\ref{Fig3}. The fit and histogram curves are vertically
    aligned by least-squares fit of the points at $C\le-0.15$
    \si{\per\nano\metre}.}\label{Sfig:LL37s}
\end{figure*}

In Fig.~\ref{Sfig:LL37s}, we show the corresponding analysis for LL-37
assuming a few alternative effective curvature sensing sites, with the
center-of-mass in the middle row. The correlation between $\theta$ and
$s$ around each peak clearly becomes more pronounced and
N-terminal-like (c.f.~Fig.~\ref{Fig5}) when the tracking site moves
towards the N-terminal end.  However, the asymmetry almost disappears
when one assumes the effective curvature sensing site to be the center
of mass of the C-terminal helix, and appears again with the opposite
C-terminal-like trend when tracking the C-terminal end. Of these
cases, the center-of-mass of the C-terminal helix is most consistent
with the symmetries of curvature tensor elements and thus acts as an
effective ``sensing site'', which indicates that this part of the
peptide is more important for curvature sensing.  Fitting the $E_C$
model to this data yields $\kappa=323\pm127\;k_BT\mathrm{nm}$,
$C_0^{-1}=-4.1\pm0.5$ nm, $b=8.2\pm 0.6$ k$_B$Tnm, and $\alpha=69\pm
2^\circ$, not significantly different from the parameters shown in
Fig.~\ref{Fig4}.

However, all distributions are still slightly asymmetric around
$s=0.5$, with average $s$-values ranging from about 0.52 to 0.51 for
the N- and C-terminal ends respectively, corresponding to an average
displacement of 0.5 nm to 0.35 nm from the mid point. A closer
examination of the significance of this observation would require
substantially better statistics, perhaps from using some enhanced
sampling method, as well as more systematic studies using a larger
range of curvatures. This is outside the scope of this study.

%% file: arxiv_main_SI.bbl
\begin{thebibliography}{65}
\providecommand{\url}[1]{\texttt{#1}}
\providecommand{\urlprefix}{ }

\bibitem[Zimmerberg and Kozlov(2006)]{Zimmerberg06}
Zimmerberg, J., and M.~M. Kozlov, 2006.
\newblock How proteins produce cellular membrane curvature.
\newblock \emph{Nat. Rev. Mol. Cell Bio.} 7:9--19.

\bibitem[Baumgart et~al.(2011)Baumgart, Capraro, Zhu, and Das]{Baumgart11}
Baumgart, T., B.~R. Capraro, C.~Zhu, and S.~L. Das, 2011.
\newblock Thermodynamics and mechanics of membrane curvature generation and
  sensing by proteins and lipids.
\newblock \emph{Annu. Rev. Phys. Chem.} 62:483--506.

\bibitem[Tonnesen et~al.(2014)Tonnesen, Christensen, Tkach, and
  Stamou]{tonnesen2014}
Tonnesen, A., S.~M. Christensen, V.~Tkach, and D.~Stamou, 2014.
\newblock Geometrical membrane curvature as an allosteric regulator of membrane
  protein structure and function.
\newblock \emph{Biophys. J.} 106:201--209.

\bibitem[Engelman(2005)]{engelman2005}
Engelman, D., 2005.
\newblock Membranes are more mosaic than fluid.
\newblock \emph{Nature} 438:578--580.

\bibitem[Fournier(1996)]{Fournier96}
Fournier, J.~B., 1996.
\newblock Nontopological saddle-splay and curvature instabilities from
  anisotropic membrane inclusions.
\newblock \emph{Phys. Rev. Lett.} 76:4436--4439.

\bibitem[Perutková et~al.(2010)Perutková, Kralj-Igli{\v c}, Frank, and
  Igli{\v c}]{Perutkova10}
Perutková, {\v S}., V.~Kralj-Igli{\v c}, M.~Frank, and A.~Igli{\v c}, 2010.
\newblock Mechanical stability of membrane nanotubular protrusions influenced
  by attachment of flexible rod-like proteins.
\newblock \emph{J. Biomech.} 43:1612--1617.

\bibitem[Ramakrishnan et~al.(2010)Ramakrishnan, Sunil~Kumar, and
  Ipsen]{ramakrishnan2010}
Ramakrishnan, N., P.~B. Sunil~Kumar, and J.~H. Ipsen, 2010.
\newblock Monte Carlo simulations of fluid vesicles with in-plane orientational
  ordering.
\newblock \emph{Phys. Rev. E} 81:041922.

\bibitem[Ramakrishnan et~al.(2011)Ramakrishnan, Kumar, and
  Ipsen]{ramakrishnan2011}
Ramakrishnan, N., P.~B.~S. Kumar, and J.~H. Ipsen, 2011.
\newblock Modeling anisotropic elasticity of fluid membranes.
\newblock \emph{Macromol. Theor. Simul.} 20:446--450.

\bibitem[Ramakrishnan et~al.(2013)Ramakrishnan, Sunil~Kumar, and
  Ipsen]{ramakrishnan2013}
Ramakrishnan, N., P.~B. Sunil~Kumar, and J.~H. Ipsen, 2013.
\newblock Membrane-mediated aggregation of curvature-inducing nematogens and
  membrane tubulation.
\newblock \emph{Biophys. J.} 104:1018--1028.

\bibitem[Akabori and Santangelo(2011)]{akabori2011}
Akabori, K., and C.~D. Santangelo, 2011.
\newblock Membrane morphology induced by anisotropic proteins.
\newblock \emph{Phys. Rev. E} 84:061909.

\bibitem[Walani et~al.(2014)Walani, Torres, and Agrawal]{walani2014}
Walani, N., J.~Torres, and A.~Agrawal, 2014.
\newblock Anisotropic spontaneous curvatures in lipid membranes.
\newblock \emph{Phys. Rev. E} 89:062715.

\bibitem[Peter et~al.(2004)Peter, Kent, Mills, Vallis, Butler, Evans, and
  McMahon]{peter2004}
Peter, B.~J., H.~M. Kent, I.~G. Mills, Y.~Vallis, P.~J.~G. Butler, P.~R. Evans,
  and H.~T. McMahon, 2004.
\newblock BAR domains as sensors of membrane curvature: the amphiphysin BAR
  structure.
\newblock \emph{Science} 303:495--499.

\bibitem[Blood and Voth(2006)]{Blood06}
Blood, P.~D., and G.~A. Voth, 2006.
\newblock Direct observation of Bin/amphiphysin/Rvs ({BAR}) domain-induced
  membrane curvature by means of molecular dynamics simulations.
\newblock \emph{Proc. Natl. Acad. Sci. U.S.A.} 103:15068--15072.

\bibitem[Drin et~al.(2007)Drin, Casella, Gautier, Boehmer, Schwartz, and
  Antonny]{drin2007}
Drin, G., J.-F. Casella, R.~Gautier, T.~Boehmer, T.~U. Schwartz, and
  B.~Antonny, 2007.
\newblock A general amphipathic $\alpha$-helical motif for sensing membrane
  curvature.
\newblock \emph{Nat. Struct. Mol. Biol.} 14:138--146.

\bibitem[Campelo et~al.(2008)Campelo, {McMahon}, and Kozlov]{campelo2008}
Campelo, F., H.~T. {McMahon}, and M.~M. Kozlov, 2008.
\newblock The hydrophobic insertion mechanism of membrane curvature
generation by proteins.
\newblock \emph{Biophys. J.} 95:2325--2339.

\bibitem[Campelo and Kozlov(2014)]{Campelo14}
Campelo, F., and M.~M. Kozlov, 2014.
\newblock Sensing membrane stresses by protein insertions.
\newblock \emph{PLoS Comput. Biol.} 10:e1003556.

\bibitem[Sodt and Pastor(2014)]{sodt2014}
Sodt, A.~J., and R.~W. Pastor, 2014.
\newblock Molecular modeling of lipid membrane curvature induction by a
  peptide: more than simply shape.
\newblock \emph{Biophys. J.} 106:1958--1969.

\bibitem[Zhu et~al.(2012)Zhu, Das, and Baumgart]{zhu2012}
Zhu, C., S.~L. Das, and T.~Baumgart, 2012.
\newblock Nonlinear sorting, curvature generation, and crowding of endophilin
  N-{BAR} on tubular membranes.
\newblock \emph{Biophys. J.} 102:1837--1845.

\bibitem[Sorre et~al.(2012)Sorre, Callan-Jones, Manzi, Goud, Prost, Bassereau,
  and Roux]{sorre2012}
Sorre, B., A.~Callan-Jones, J.~Manzi, B.~Goud, J.~Prost, P.~Bassereau, and
  A.~Roux, 2012.
\newblock Nature of curvature coupling of amphiphysin with membranes depends on
  its bound density.
\newblock \emph{Proc. Natl. Acad. Sci. U.S.A.} 109:173--178.

\bibitem[Aimon et~al.(2014)Aimon, Callan-Jones, Berthaud, Pinot, Toombes, and
  Bassereau]{Aimon14}
Aimon, S., A.~Callan-Jones, A.~Berthaud, M.~Pinot, G.~E.~S. Toombes, and
  P.~Bassereau, 2014.
\newblock Membrane shape modulates transmembrane protein distribution.
\newblock \emph{Dev. Cell} 28:212--218.

\bibitem[Shi and Baumgart(2015)]{shi2015}
Shi, Z., and T.~Baumgart, 2015.
\newblock Membrane tension and peripheral protein density mediate membrane
  shape transitions.
\newblock \emph{Nat. Commun.} 6:5974.

\bibitem[Hsieh et~al.(2012)Hsieh, Hsu, Capraro, Wu, Chen, Yang, and
  Baumgart]{hsieh2012}
Hsieh, W.-T., C.-J. Hsu, B.~R. Capraro, T.~Wu, C.-M. Chen, S.~Yang, and
  T.~Baumgart, 2012.
\newblock Curvature sorting of peripheral proteins on solid-supported wavy
  membranes.
\newblock \emph{Langmuir} 28:12838--12843.

\bibitem[Ramesh et~al.(2013)Ramesh, Baroji, Reihani, Stamou, Oddershede, and
  Bendix]{ramesh2013}
Ramesh, P., Y.~F. Baroji, S.~N.~S. Reihani, D.~Stamou, L.~B. Oddershede, and
  P.~M. Bendix, 2013.
\newblock {FBAR} syndapin 1 recognizes and stabilizes highly curved tubular
  membranes in a concentration dependent manner.
\newblock \emph{Sci. Rep.} 3:1565.

\bibitem[Black et~al.(2014)Black, Cheney, Campbell, and Knowles]{black2014}
Black, J.~C., P.~P. Cheney, T.~Campbell, and M.~K. Knowles, 2014.
\newblock Membrane curvature based lipid sorting using a nanoparticle patterned
  substrate.
\newblock \emph{Soft Matter} 10:2016.

\bibitem[Zasloff(1987)]{zasloff1987}
Zasloff, M., 1987.
\newblock Magainins, a class of antimicrobial peptides from Xenopus skin:
  isolation, characterization of two active forms, and partial {cDNA} sequence
  of a precursor.
\newblock \emph{Proc. Natl. Acad. Sci. U.S.A.} 84:5449--5453.

\bibitem[Habermann(1972)]{habermann1972}
Habermann, E., 1972.
\newblock Bee and wasp venoms.
\newblock \emph{Science} 177:314--322.

\bibitem[Gudmundsson et~al.(1996)Gudmundsson, Agerberth, Odeberg, Bergman,
  Olsson, and Salcedo]{gudmundsson1996}
Gudmundsson, G.~H., B.~Agerberth, J.~Odeberg, T.~Bergman, B.~Olsson, and
  R.~Salcedo, 1996.
\newblock The human gene {FALL}39 and processing of the cathelin precursor to
  the antibacterial peptide {LL}-37 in granulocytes.
\newblock \emph{Eur. J. Biochem.} 238:325--332.

\bibitem[Melo et~al.(2009)Melo, Ferre, and Castanho]{Melo09}
Melo, M.~N., R.~Ferre, and M.~A. R.~B. Castanho, 2009.
\newblock Antimicrobial peptides: linking partition, activity and high
  membrane-bound concentrations.
\newblock \emph{Nat. Rev. Microbiol.} 7:245--250.

\bibitem[Ludtke et~al.(1996)Ludtke, He, Heller, Harroun, Yang, and
  Huang]{ludtke1996}
Ludtke, S.~J., K.~He, W.~T. Heller, T.~A. Harroun, L.~Yang, and H.~W. Huang,
  1996.
\newblock Membrane Pores Induced by Magainin†.
\newblock \emph{Biochemistry} 35:13723--13728.

\bibitem[Yang et~al.(2001)Yang, Harroun, Weiss, Ding, and Huang]{yang2001}
Yang, L., T.~A. Harroun, T.~M. Weiss, L.~Ding, and H.~W. Huang, 2001.
\newblock Barrel-stave model or toroidal model? {A} case study on melittin
  pores.
\newblock \emph{Biophys. J.} 81:1475--1485.

\bibitem[Leontiadou et~al.(2006)Leontiadou, Mark, and Marrink]{leontiadou2006}
Leontiadou, H., A.~E. Mark, and S.~J. Marrink, 2006.
\newblock Antimicrobial peptides in action.
\newblock \emph{J. Am. Chem. Soc.} 128:12156--12161.

\bibitem[Henzler~Wildman et~al.(2003)Henzler~Wildman, Lee, and
  Ramamoorthy]{henzler2003}
Henzler~Wildman, K.~A., D.-K. Lee, and A.~Ramamoorthy, 2003.
\newblock Mechanism of lipid bilayer disruption by the human antimicrobial
  peptide, {LL}-37.
\newblock \emph{Biochemistry} 42:6545--6558.

\bibitem[Anthony~G.(2011)]{lee2011}
Anthony~G., L., 2011.
\newblock Biological membranes: the importance of molecular detail.
\newblock \emph{Trends in Biochemical Sciences} 36:493--500.

\bibitem[Sun et~al.(2015)Sun, Forsman, and Woodward]{sun2015}
Sun, D., J.~Forsman, and C.~E. Woodward, 2015.
\newblock Amphipathic membrane-active peptides recognize and stabilize ruptured
  membrane pores: exploring cause and effect with coarse-grained simulations.
\newblock \emph{Langmuir} 31:752--761.

\bibitem[Wang et~al.(2014)Wang, Mishra, Epand, and Epand]{wang2014}
Wang, G., B.~Mishra, R.~F. Epand, and R.~M. Epand, 2014.
\newblock High-quality 3D structures shine light on antibacterial, anti-biofilm
  and antiviral activities of human cathelicidin {LL}-37 and its fragments.
\newblock \emph{BBA - Biomembranes} 1838:2160--2172.

\bibitem[Noguchi(2011)]{noguchi2011}
Noguchi, H., 2011.
\newblock Anisotropic surface tension of buckled fluid membranes.
\newblock \emph{Phys. Rev. E} 83:061919.

\bibitem[Hu et~al.(2013)Hu, Diggins, and Deserno]{Hu2013}
Hu, M., P.~Diggins, and M.~Deserno, 2013.
\newblock Determining the bending modulus of a lipid membrane by simulating
  buckling.
\newblock \emph{J. Chem. Phys.} 138:214110--214110--13.

\bibitem[Cui et~al.(2011)Cui, Lyman, and Voth]{cui2011}
Cui, H., E.~Lyman, and G.~A. Voth, 2011.
\newblock Mechanism of membrane curvature sensing by amphipathic helix
  containing proteins.
\newblock \emph{Biophys. J.} 100:1271--1279.

\bibitem[Wang(2008)]{wang2008}
Wang, G., 2008.
\newblock Structures of human host defense cathelicidin {LL}-37 and its
  smallest antimicrobial peptide {KR}-12 in lipid micelles.
\newblock \emph{J. Biol. Chem.} 283:32637--32643.
\newblock PDB: 2K6O.

\bibitem[Hara et~al.(2001)Hara, Kodama, Kondo, Wakamatsu, Takeda, Tachi, and
  Matsuzaki]{hara2001}
Hara, T., H.~Kodama, M.~Kondo, K.~Wakamatsu, A.~Takeda, T.~Tachi, and
  K.~Matsuzaki, 2001.
\newblock Effects of peptide dimerization on pore formation: Antiparallel
  disulfide-dimerized magainin 2 analogue.
\newblock \emph{Biopolymers} 58:437--446.
\newblock {PDB: 1DUM}.

\bibitem[Terwilliger et~al.(1982)Terwilliger, Weissman, and
  Eisenberg]{terwilliger1982}
Terwilliger, T.~C., L.~Weissman, and D.~Eisenberg, 1982.
\newblock The structure of melittin in the form {I} crystals and its
  implication for melittin's lytic and surface activities.
\newblock \emph{Biophys. J.} 37:353--361.
\newblock PDB: 2MLT.

\bibitem[Marrink et~al.(2007)Marrink, Risselada, Yefimov, Tieleman, and
  de~Vries]{Marrink07}
Marrink, S.~J., H.~J. Risselada, S.~Yefimov, D.~P. Tieleman, and A.~H.
  de~Vries, 2007.
\newblock The {MARTINI} force field:  coarse grained model for biomolecular
  simulations.
\newblock \emph{J. Phys. Chem. B} 111:7812--7824.

\bibitem[Hatzakis et~al.(2009{\natexlab{a}})Hatzakis, Bhatia, Larsen, Madsen,
  Bolinger, Kunding, Castillo, Gether, Hedegård, and Stamou]{Hatzakis2009}
Hatzakis, N.~S., V.~K. Bhatia, J.~Larsen, K.~L. Madsen, P.~Bolinger, A.~H.
  Kunding, J.~Castillo, U.~Gether, P.~Hedegård, and D.~Stamou, 2009.
\newblock How curved membranes recruit amphipathic helices and protein
  anchoring motifs.
\newblock \emph{Nat. Chem. Biol.} 5:835--841.

\bibitem[Vanni et~al.(2014)Vanni, Hirose, Barelli, Antonny, and
  Gautier]{vanni2014}
Vanni, S., H.~Hirose, H.~Barelli, B.~Antonny, and R.~Gautier, 2014.
\newblock A sub-nanometre view of how membrane curvature and composition
  modulate lipid packing and protein recruitment.
\newblock \emph{Nat. Commun.} 5:4916.

\bibitem[Pronk et~al.(2013)Pronk, Páll, Schulz, Larsson, Bjelkmar, Apostolov,
  Shirts, Smith, Kasson, van~der Spoel, Hess, and Lindahl]{Pronk13}
Pronk, S., S.~Páll, R.~Schulz, P.~Larsson, P.~Bjelkmar, R.~Apostolov, M.~R.
  Shirts, J.~C. Smith, P.~M. Kasson, D.~van~der Spoel, B.~Hess, and E.~Lindahl,
  2013.
\newblock {GROMACS} 4.5: a high-throughput and highly parallel open source
  molecular simulation toolkit.
\newblock \emph{Bioinformatics} 29:845--854.

\bibitem[Monticelli et~al.(2008)Monticelli, Kandasamy, Periole, Larson,
  Tieleman, and Marrink]{monticelli2008}
Monticelli, L., S.~K. Kandasamy, X.~Periole, R.~G. Larson, D.~P. Tieleman, and
  S.-J. Marrink, 2008.
\newblock The {MARTINI} coarse-grained force field: extension to proteins.
\newblock \emph{J. Chem. Theory Comput.} 4:819--834.

\bibitem[Yesylevskyy et~al.(2010)Yesylevskyy, Sch{\"a}fer, Sengupta, and
  Marrink]{yesylevskyy2010}
Yesylevskyy, S.~O., L.~V. Sch{\"a}fer, D.~Sengupta, and S.~J. Marrink, 2010.
\newblock Polarizable water model for the coarse-grained {MARTINI} force
  field.
\newblock \emph{PLoS Comput. Biol.} 6:e1000810.

\bibitem[Marrink et~al.(2004)Marrink, de~Vries, and Mark]{Marrink04}
Marrink, S.~J., A.~H. de~Vries, and A.~E. Mark, 2004.
\newblock Coarse grained model for semiquantitative lipid simulations.
\newblock \emph{J. Phys. Chem. B} 108:750--760.

\bibitem[Baoukina et~al.(2007)Baoukina, Monticelli, Amrein, and
  Tieleman]{Baoukina07}
Baoukina, S., L.~Monticelli, M.~Amrein, and D.~P. Tieleman, 2007.
\newblock The molecular mechanism of monolayer-bilayer transformations of lung
  surfactant from molecular dynamics simulations.
\newblock \emph{Biophys. J.} 93:3775--3782.

\bibitem[de~Jong et~al.(2013)de~Jong, Singh, Bennett, Arnarez, Wassenaar,
  Schäfer, Periole, Tieleman, and Marrink]{dejong2013}
de~Jong, D.~H., G.~Singh, W.~F.~D. Bennett, C.~Arnarez, T.~A. Wassenaar, L.~V.
  Schäfer, X.~Periole, D.~P. Tieleman, and S.~J. Marrink, 2013.
\newblock Improved parameters for the {Martini} coarse-grained protein force
  field.
\newblock \emph{J. Chem. Theory Comput.} 9:687--697.

\bibitem[Bussi et~al.(2007)Bussi, Donadio, and Parrinello]{Bussi07}
Bussi, G., D.~Donadio, and M.~Parrinello, 2007.
\newblock Canonical sampling through velocity rescaling.
\newblock \emph{J. Chem. Phys.} 126:014101.

\bibitem[Berendsen et~al.(1984)Berendsen, Postma, van Gunsteren, DiNola, and
  Haak]{berendsen1984}
Berendsen, H. J.~C., J.~P.~M. Postma, W.~F. van Gunsteren, A.~DiNola, and J.~R.
  Haak, 1984.
\newblock Molecular dynamics with coupling to an external bath.
\newblock \emph{J. Chem. Phys.} 81:3684--3690.

\bibitem[Essmann et~al.(1995)Essmann, Perera, Berkowitz, Darden, Lee, and
  Pedersen]{essmann1995}
Essmann, U., L.~Perera, M.~L. Berkowitz, T.~Darden, H.~Lee, and L.~G. Pedersen,
  1995.
\newblock A smooth particle mesh Ewald method.
\newblock \emph{J. Chem. Phys.} 103:8577--8593.

\bibitem[Lee et~al.(2013)Lee, Sun, Hung, and Huang]{lee2013}
Lee, M.-T., T.-L. Sun, W.-C. Hung, and H.~W. Huang, 2013.
\newblock Process of inducing pores in membranes by melittin.
\newblock \emph{Proc. Natl. Acad. Sci. U.S.A.} 110:14243--14248.

\bibitem[Kreyszig(1991)]{kreyszig1991}
Kreyszig, E., 1991.
\newblock Differential geometry.
\newblock Dover Publications, New York.

\bibitem[K{\"u}nsch(1989)]{kunsch1989}
K{\"u}nsch, H.~R., 1989.
\newblock The Jackknife and the Bootstrap for general stationary observations.
\newblock \emph{Ann. Stat.} 17:1217--1241.

\bibitem[Humphrey et~al.(1996)Humphrey, Dalke, and Schulten]{vmd96}
Humphrey, W., A.~Dalke, and K.~Schulten, 1996.
\newblock {VMD}: Visual molecular dynamics.
\newblock \emph{J. Mol. Graphics} 14:33--38.

\bibitem[Hatzakis et~al.(2009{\natexlab{b}})Hatzakis, Bhatia, Larsen, Madsen,
  Bolinger, Kunding, Castillo, Gether, Hedeg{\aa}rd, and Stamou]{Hatzakis09}
Hatzakis, N.~S., V.~K. Bhatia, J.~Larsen, K.~L. Madsen, P.-Y. Bolinger, A.~H.
  Kunding, J.~Castillo, U.~Gether, P.~Hedeg{\aa}rd, and D.~Stamou, 2009.
\newblock How curved membranes recruit amphipathic helices and protein
  anchoring motifs.
\newblock \emph{Nat. Chem. Biol.} 5:835--841.

\bibitem[He and Lazaridis(2013)]{he2013}
He, Y., and T.~Lazaridis, 2013.
\newblock Activity determinants of helical antimicrobial peptides: A
  large-scale computational study.
\newblock \emph{PLoS ONE} 8:e66440.

\bibitem[Braun et~al.(2012)Braun, Sevcsik, Chin, Rhoades, Tristram-Nagle, and
  Sachs]{Braun12}
Braun, A.~R., E.~Sevcsik, P.~Chin, E.~Rhoades, S.~Tristram-Nagle, and J.~N.
  Sachs, 2012.
\newblock $\alpha$-Synuclein induces both positive mean curvature and negative
  Gaussian curvature in membranes.
\newblock \emph{J. Am. Chem. Soc.} 134:2613--2620.

\bibitem[Goodsell and Olson(2000)]{goodsell00}
Goodsell, D.~S., and A.~J. Olson, 2000.
\newblock Structural symmetry and protein function.
\newblock \emph{Annu. Rev. Biophys. Biomol. Struct.} 29:105--153.

\bibitem[Rosenberg et~al.(2005)Rosenberg, Quinlan, Forkey, and
  Goldman]{rosenberg2005}
Rosenberg, S.~A., M.~E. Quinlan, J.~N. Forkey, and Y.~E. Goldman, 2005.
\newblock Rotational motions of macromolecules by single-molecule
  fluorescence microscopy.
\newblock \emph{Acc. Chem. Res.} 38:583--593.

\bibitem[Davies et~al.(2012)Davies, Anselmi, Wittig, Faraldo-Gómez, and
  Kühlbrandt]{davies2012}
Davies, K.~M., C.~Anselmi, I.~Wittig, J.~D. Faraldo-Gómez, and W.~Kühlbrandt,
  2012.
\newblock Structure of the yeast {F1Fo-ATP} synthase dimer and its role in
  shaping the mitochondrial cristae.
\newblock \emph{Proc. Natl. Acad. Sci. U.S.A.} 19:13602--13607.

\bibitem[Mim et~al.(2012)Mim, Cui, Gawronski-Salerno, Frost, Lyman, Voth, and
  Unger]{mim2012}
Mim, C., H.~Cui, J.~A. Gawronski-Salerno, A.~Frost, E.~Lyman, G.~A. Voth, and
  V.~M. Unger, 2012.
\newblock Structural basis of membrane bending by the {N-BAR} protein
  endophilin.
\newblock \emph{Cell} 149:137--145.

\bibitem[Ge et~al.(2014)Ge, G{\'o}mez-Llobregat, Skwark, Ruysschaert,
  Wieslander, and Lind{\'e}n]{ge2014}
Ge, C., J.~G{\'o}mez-Llobregat, M.~J. Skwark, J.-M. Ruysschaert, {\r
  A}.~Wieslander, and M.~Lind{\'e}n, 2014.
\newblock Membrane remodeling capacity of a vesicle-inducing
  glycosyltransferase.
\newblock \emph{{FEBS} J.} 281:3667--3684.
\end{thebibliography}
